\documentclass[12pt,epsf]{article}
\usepackage{epsfig}

\renewcommand{\thanks}[1]{\footnote{#1}} % Use this for footnotes

\newcommand{\be}{\begin{equation}}
\newcommand{\ee}{\end{equation}}
\newcommand{\bea}{\begin{eqnarray}}
\newcommand{\eea}{\end{eqnarray}}

\begin{document}

\pagestyle{empty}

\bigskip\bigskip
\begin{center}
{\bf \large Diagonal Forms of a Dual Scale Cosmology}
\end{center}

\begin{center}
James Lindesay\footnote{e-mail address, jlslac@slac.stanford.edu} \\
Computational Physics Laboratory \\
Howard University,
Washington, D.C. 20059 
\end{center}

\begin{center}
{\bf Abstract}
\end{center}

A hybrid metric with off-diagonal temporal-radial behavior
that was constructed to conveniently parameterized
the early and late time behaviors of the universe is shown
to have diagonal forms consistent with Robertson-Walker
and deSitter geometries.  The dynamics of the energy
content of the cosmology as parameterized by the
classical thermal fraction is briefly discussed as
motivation for the comparison of the observables predicted
by various micoscopic models of the early evolution
of the universe.

\setcounter{equation}{0}
\section{Introduction}
\indent

There has been a considerable amount of interest in
developing convenient frameworks for exploring the
dynamics of the very early universe. 
In previous work\cite{FRWdS, ThermalDual} a metric framework
was developed that explicitly demonstrated the evolution of dual
scales that could be associated with the microscopic
and macroscopic behaviors of the cosmology. 
The temporal dynamics of these
otherwise independent scales were shown to
be connected only through the Einstein equation
(in the absence of a cosmological constant).
The metric, given by
\be
\begin{array}{r}
ds^2 =
 -  c^2 dt^2 + R^2(ct) \left( dr - {r \over R_v (ct)} c dt  \right )^2  \\
+ R^2 (ct) \left ( r^2 d \theta ^2 + r^2 sin^2 \theta d\phi ^2   \right ) ,
\end{array}
\label{FRWdSmetric}
\ee
constrains the dynamics of an ideal fluid as expressed by
\be
\begin{array}{l}
{d \over d ct} \rho = -\sqrt{{24 \pi G_N \rho  \over c^4 }} (P + \rho) \\ \\
\rho= {3 c^4 \over 8 \pi G_N} \left (
 {\dot{R} \over R } + {1 \over R_v}    \right )^2 .
\label{rhodynamics}
\end{array}
\ee
The energy content is assumed to be of the form
$\rho=\rho_v + \rho_{thermal}$ in terms of gravitational
vacuum/condensate energy and thermal/classical energy,
and the equations of state for the various components
take the usual form $P_s=w_s \rho_s$.

If one defines the thermal (classical) fraction by
$f(ct) \equiv {\rho_{thermal} \over \rho}$ and the scale $R_v$ is given
by $\sqrt{{8 \pi G_N \rho_v \over 3 c^4}} = {1 \over R_v}$, the evolution
equation satisfies
\be
\begin{array}{l}
{d \over d ct} \rho = -  \left [ (1 + w_v) 
+(w-w_v) f(ct) \right ] \rho
\sqrt{{24 \pi G_N  \over c^4 } \rho } , \\ \\
\left ( {\rho_* \over \rho} \right )^{1/2}= 1 + {3 \over 2}
\sqrt{{8 \pi G_N \rho_* \over 3 c^4}} \int _{ct_*} ^{ct}  \left [ (1 + w_v) 
+(w-w_v) f(ct') \right ] c dt'.
\end{array}
\label{rhosolution}
\ee
The energy density and condensate scale are expected to have extreme
values given by $\rho_I \Leftarrow \rho \Rightarrow \rho_\Lambda$
and $R_I \Leftarrow R_v \Rightarrow R_\Lambda$
for $0 \leftarrow ct \rightarrow \infty$.
The initial and final states are
taken to have a vanishing thermal fraction.  The solution
Eq. \ref{rhosolution} can then be used to determine the temporal behavior
of the thermal scale\cite{ThermalDual} $R$ using Eq. \ref{rhodynamics}.

\setcounter{equation}{0}
\section{Metric Diagonal Forms}
\indent

For many, coordinates for which the metric form is diagonal
provide the most intuitive construct of a given geometry. 
Therefore, diagonal forms will be constructed for the metric
form in Eq. \ref{FRWdSmetric} using coordinate transformations
on the coordinates $(ct, r)$.  Two coordinate forms will be
explored; one directly relating the dual scales to that of the
Robertson-Walker (RW) geometry\cite{RWgeometry}, and the other analogous
to a deSitter geometry\cite{deSittergeometry}.

\subsection{Radial coordinate transformation}
\indent

A transformation on the radial coordinate will be sought
to diagonalize the form in Eq. \ref{FRWdSmetric}.  Using
Eq. \ref{rhodynamics}, one is motivated to define the
reduced scale factor $\mathcal{R}$, and require angular
isotropy of the metric expressed in either coordinate
system:
\be
{\dot{\mathcal{R}} \over \mathcal{R}} \equiv {\dot{R} \over R}
+ {1 \over R_v} \quad , \quad
R \, r = \mathcal{R} \, r_{RW} \quad .
\ee
A brief and straightforward calculation yields
\be
R \, dr = \mathcal{R} \left [ {r_{RW} \over R_v} dct +
dr_{RW}    \right ] \, .
\ee
Substitution into Eq. \ref{FRWdSmetric} gives
\be
ds^2 = -c^2 dt^2 + \mathcal{R}^2(ct)  \left [ dr_{RW} ^2 +
 r_{RW}^2 d \theta ^2 + r_{RW} ^2 sin^2 \theta \, d\phi ^2   \right ].
\ee
This demonstrates that the reduced scale parameter
$\mathcal{R}$ for coordinates $(ct,r,\theta,\phi)$
corresponds to the standard Robertson-Walker scale parameter for coordinates
$(ct, r_{RW},\theta, \phi)$.

\subsection{Temporal coordinate transformation}
\indent

Next, a coordinate transformation on the temporal coordinate
will be sought to diagonalize the metric form.  For conciseness,
the following forms will be defined:
\be
\Delta(ct) \equiv {R(ct) \over R_v (ct)} \quad , \quad
d\tilde{t} \equiv {c dt \over R(ct)} \quad .
\ee
Rewriting the metric in Eq. \ref{FRWdSmetric}
\be
ds^2 = R^2 \left [
-d \tilde{t} ^2 + (dr - r \Delta d\tilde{t})^2 +
 r^2 d \theta ^2 + r ^2 sin^2 \theta \, d\phi ^2
\right ] ,
\ee
all dimension is carried in the parameter $R(ct)$.
The metric becomes diagonal under the
temporal transformation
\be
\begin{array}{l}
d \tilde{t} = B(ct_{dS},r) \, c dt_{dS} -
{r \Delta \over 1 - (r \Delta)^2} \, dr  = \\ \\
{c dt \over R(ct)} = B(ct_{dS},r) \, c dt_{dS} + {1 \over 2 \Delta} 
{\partial \over \partial r}
log( 1 - (r \Delta)^2)  \,dr ,
\end{array}
\ee
giving a metric of the form
\be
ds^2 = R^2 \left ( - \left [ 1 - (r \Delta)^2 
 \right ] B^2 c^2 dt_{dS}^2 +
{dr^2 \over \left [ 1 - (r \Delta)^2  \right ] }+
 r^2 d \theta ^2 + r ^2 sin^2 \theta \, d\phi ^2
\right ).
\label{deSittertime}
\ee
Thus, there is a space-time coordinate singularity at
$r \Delta(ct) =1$ corresponding to a
coordinate horizon (if $B$ is not singular
on that surface).  The integrability condition
for functions of the transformed
coordinates constrains the form of the
function $B$.  A general form is given by
\be
\begin{array}{l}
B(ct_{dS},r) =b(ct_{dS}) + 
{\partial \over \partial c t_{dS}} \left [
{1 \over 2 \Delta} log( 1 - (r \Delta)^2) \right ] \\
\quad \quad \quad = b(ct_{dS}) + B(ct_{dS},r) R
{\partial \over \partial c t} \left [
{1 \over 2 \Delta} log( 1 - (r \Delta)^2) \right ] .
\end{array}
\label{Beqn}
\ee
In the cosmology of interest, the scales satisfy
dynamics in a manner that the form
$B$ will be non-singular.  The metric in
Eq. \ref{deSittertime} corresponds to the usual deSitter coordinates
$(ct_{dS} R,r R,\theta, \phi)$ with horizon scale $R_v$ if
the scale parameters $R$ and $R_v$ are
constants and $b=1$.

\setcounter{equation}{0}
\section{Conclusions}
\indent

It has been demonstrated that the geometry described by
the dual scale metric of interest is of the Robertson-Walker
form without a cosmological constant,
where the RW radial scale factor decomposes into
components that are convenient for exploring the
various extremum epochs of the cosmology.  The
early and late stages of the cosmology are conveniently
described by macroscopic densities $\rho_v$ and scales
$R_v$ using geometries similar to that of deSitter,
while the intermediate behavior is essentially that of
a classical Robertson-Walker geometry driven by
thermal energy densities.  Future work will utilize
the transition behavior of this metric in manners similar
to those in the references\cite{NSBP06, ThermalDual}
in an attempt to microscopically characterize the
early form of the cosmology consistent with
intermediate astrophysical observations.

\end{document}